\documentclass{cpbtex} 

\usepackage{graphicx}  
\usepackage{float}
\usepackage{bm}        
\usepackage{slashed}   
\usepackage{color} 
\usepackage{subfigure}
\usepackage{epstopdf}
\usepackage[numbers,sort&compress,mcite]{natbib}
\usepackage{hyperref}
\usepackage{verbatim}

\graphicspath{{figure/}}

\newcounter{YJC}

\graphicspath{{Figures/}}

\begin{document}

\title{Study of the gluonic quartic gauge couplings at muon colliders}

\author{Ji-Chong Yang$^{1,2}$\thanks{E-mail:yangjichong@lnnu.edu.cn}, Yu-Chen Guo$^{1,2}$\thanks{E-mail:ycguo@lnnu.edu.cn, Corresponding author.}, \ and \ \ Yi-Fei Dong$^{1}$\thanks{E-mail:dyf2818051165@163.com}\\
$^{1}${Department of Physics, Liaoning Normal University, Dalian 116029, China}\\
$^{2}${Center for Theoretical and Experimental High Energy Physics,} \\
{Liaoning Normal University, Dalian 116029, China}}

\date{\today}

\maketitle
\begin{abstract}
The potential of the muon colliders open up new possibilities for the exploration of new physics beyond the Standard Model.
It is worthwhile to investigate whether muon colliders are suitable for studying gluonic quartic gauge couplings~(gQGCs), which can be contributed by dimension-8 operators in the framework of the Standard Model effective field theory, and are intensively studied recently.
In this paper, we study the sensitivity of the process $\mu^+\mu^-\to j j \nu\bar{\nu}$ to gQGCs.
Our result indicate that the muon colliders with c.m. energies larger than $4\;{\rm TeV}$ can be more sensitive to gQGCs than the Large Hadron Collider.
\end{abstract}



\section{\label{sec1}Introduction}

So far, most of experimental results have demonstrated the success of the Standard Model~(SM).
However, the theoretical difficulties of the theory itself have led people to believe that the SM is merely an effective theory, with the expectation of new physics (NP) potentially emerging at higher energy scales~\cite{johnellis}.
Due to the lack of clear sign of new physics, the SM effective field theory (SMEFT)~\cite{weinberg,SMEFTReview1,SMEFTReview2,SMEFTReview3} has become a popular theoretical tool for studying experimental results in a model-independent way.
In most cases, NP effects are expected to manifest in physical processes that are described by dimension-6 operators of SMEFT.
However, in some special cases, the contribution of dimension-8 operators may be more important~\cite{BI,Ellis:2018cos,d81,looportree,aqgcold,aqgcnew}.
These cases include anomalous quartic gauge coupling (aQGC) and neutral triple gauge coupling (nTGC), which have been the subject of many theoretical studies and experimental measurements \cite{sswwexp1,sswwexp2,zaexp1,zaexp2,zaexp3,waexp1,zzexp1,zzexp2,wzexp1,wzexp2,wwexp1,wwexp2,wvzvexp,waexp2,zzexp3,ntgc1,ntgc2,ntgc3,ntgc4,ntgc5,ntgc6,ntgc7,wwstudy,wastudy,zastudy,wwwwunitary,Yang:2022ilt,Yang:2022fhw,triphoton,Zhang:2023yfg,Ellis:2022zdw}.
These couplings are fixed by electroweak gauge bosons only.
Another type of anomalous gauge coupling that does not exist in the SM is the quartic coupling between gluons and electroweak~(EW) vector bosons.
They can be described by dimension-8 operators, which are operators contributing to gluonic quartic gauge couplings~(gQGCs) \cite{Ellis:2018cos}.
The quartic couplings of gluons to EW gauge bosons arise in the Born-Infeld extension of the SM~(BI), which was originally proposed to set an upper limit on the strength of the electromagnetic field~\cite{BI}.
It has been shown that the BI also appears in the theories inspired by M theory \cite{Bachas:1995kx,Fradkin:1985qd,Tseytlin:1999dj}.
Recently, the gQGCs have been studied at the Large Hadron Collider~(LHC)~\cite{Ellis:2018cos,Ellis:2021dfa}.

High-energy collisions can directly generate new particles, while precise measurements assist us in identifying the indirect effects of unknown new physics and understanding the dynamics of known particles.
The muon collider capitalizes on the advantages of two strategic approaches to exploit the complementarity between energy and precision.
The inception of concepts pertaining to muon colliders dates back to earlier time~\cite{Tikhonin:2008pw,Neuffer:1979gq,Cline:1980sa,Skrinsky:1981ht,Neuffer:1983jr}.
At present, the energy frontier attainable by muon colliders remains undetermined.
Ongoing investigations are centered on a 10 TeV configuration, with a targeted integrated luminosity of 10 ab$^{-1}$~\cite{Black:2022cth,Accettura:2023ked}.
With the high-energy and high-luminosity, the muon collider not only possesses a stronger capability for probing NP compared to the LHC, but also has the ability for more precisely measures.
As a result, high energy muon colliders have gained much attention in the community~\cite{Han:2020pif,Han:2020uak,Liu:2021jyc,Liu:2021akf,Han:2021udl,Han:2021lnp,Han:2022mzp,Han:2022ubw,Han:2022edd,Yang:2022ilt,Yang:2022fhw,triphoton,Zhang:2023yfg,Chowdhury:2023imd,Jueid:2023zxx}.
The muon colliders are ideal places to study the dimension-8 operators, including those contributing to the gQGCs.
In a high-energy muon collider, the initial state muons emit low-virtuality vector bosons.
The weak-boson fusion process transforms the muon collider into a high-luminosity vector boson collider \cite{Aime:2022flm,Black:2022cth}.
The gQGCs can affect the process $\mu^+\mu^-\to j j \nu\bar{\nu}$ via both the vector boson fusion (VBF) processes and the tri-boson productions.
In this paper, we study the sensitivity of the process $\mu^+\mu^-\to j j \nu\bar{\nu}$ to the gQGCs.

The rest of this paper is organized as follows.
In section \ref{sec:gQGC:operators}, the dimension-8 operators contributing to gQGCs are introduced.
In section \ref{sec3}, we compare the VBF process with the tri-boson process, and discuss unitarity bound on the operator coefficients.
We also present our event selection strategy at muon colliders with different energies and luminosities.
The numerical results of the constraints on the coefficients are presented in section \ref{sec4}.
In section \ref{sec5}, we summarize and draw our conclusions.

\section{Dimension-8 Gluonic QGC Operators}
\label{sec:gQGC:operators}

Although there are many possibilities for new physics at higher energy scales above the EW one, the low energy effective field theory (EFT) should be subject to the SM SU(3)$_c$ $\times $SU(2)$_L \times $U(1)$_Y$ gauge symmetries.
A convenient way to take these symmetries into account is the SMEFT \cite{Buchmuller:1985jz}, which includes systematically all the allowed interactions with mass dimension $d > 4$.
The extra dimensions are compensated by inverse powers of a mass scale $M$ that is associated with heavy new particles.
The gQGCs appear at dimension-8 level with $1/M^4$ suppression \cite{Ellis:2018cos},

\begin{subequations}
\label{basis}
\begin{eqnarray}
  O_{gT,0}
& \equiv &
  \frac 1 {16 M^4_0}
       \sum_a G^a_{\mu \nu} G^{a, \mu \nu}
\times \sum_i W^i_{\alpha \beta} W^{i, \alpha \beta} \,,
\\
  O_{gT,1}
& \equiv &
  \frac 1 {16 M^4_1}
       \sum_a G^a_{\alpha \nu} G^{a, \mu \beta}
\times \sum_i W^i_{\mu \beta} W^{i, \alpha \nu} \,,
\\
  O_{gT,2}
& \equiv &
  \frac 1 {16 M^4_2}
       \sum_a G^a_{\alpha \mu} G^{a, \mu \beta}
\times \sum_i W^i_{\nu \beta} W^{i, \alpha \nu} \,,
\\
  O_{gT,3}
& \equiv &
  \frac 1 {16 M^4_3}
       \sum_a G^a_{\alpha \mu} G^a_{\beta \nu}
\times \sum_i W^{i, \mu \beta} W^{i, \nu \alpha} \,,
\\
  O_{gT,4}
& \equiv &
  \frac 1 {16 M^4_4}
       \sum_a G^a_{\mu \nu} G^{a, \mu \nu}
\times B_{\alpha \beta} B^{\alpha \beta} \,,
\\
  O_{gT,5}
& \equiv &
  \frac 1 {16 M^4_5}
       \sum_a G^a_{\alpha \nu} G^{a, \mu \beta}
\times B_{\mu \beta} B^{\alpha \nu} \,,
\\
  O_{gT,6}
& \equiv &
  \frac 1 {16 M^4_6}
       \sum_a G^a_{\alpha \mu} G^{a, \mu \beta}
\times B_{\nu \beta} B^{\alpha \nu} \,,
\\
  O_{gT,7}
& \equiv &
  \frac 1 {16 M^4_7}
       \sum_a G^a_{\alpha \mu} G^a_{\beta \nu}
\times B^{\mu \beta} B^{\nu \alpha} \,,
\end{eqnarray}
\end{subequations}
%
where $G^a_{\mu \nu}$ is gluon field strengths, $W^i_{\mu \nu}$ and $B_{\mu \nu}$ are electroweak field strengths.
Since gluons carry QCD color, denoted by the $a$ superscript of $G^a_{\mu \nu}$, the gQGC operators must contain an even number of gluon field strengths, such as $G^a_{\mu \nu} G^{a, \alpha \beta}$, so as to be colorless.
The same thing applies for the SU(2)$_L \times $U(1)$_Y$ gauge boson field strengths, for example $W^i_{\mu \nu} W^{i, \alpha \beta}$ and $B_{\mu \nu} B^{\alpha \beta}$.
Another symmetry to be imposed is Lorentz invariance.
There are four different Lorentz-invariant contractions, as shown above.
Hence we must consider eight gQGC operators in total.

The total and differential cross-sections are largely determined by the Lorentz structure of the gQGC operators.
The eight operators can be classified into four pairs
\{$ O_{gT,(0,4)}$,
$ O_{gT,(1,5)}$,
$ O_{gT,(2,6)}$,
$ O_{gT,(3,7)}$\},
each with a same Lorentz structure.

The hierarchical structure of the cross-sections generated by the eight gQGC operators is manifest in the $95\%$ C.L. lower bounds derived from the ATLAS data~\cite{Ellis:2018cos} are
\begin{equation}
\begin{split}
M_0\geq 1040\;{\rm GeV},\;&M_1\geq 777\;{\rm GeV},\\
M_2\geq 750\;{\rm GeV},\;&M_3\geq 709\;{\rm GeV},\\
M_4\geq 1399\;{\rm GeV},\;&M_5\geq 1046\;{\rm GeV},\\
M_6\geq 1010\;{\rm GeV},\;&M_7\geq 954\;{\rm GeV}.
\end{split}
\label{eq.1.1}
\end{equation}

\section{\label{sec3}Features of the signal}

\subsection{\label{sec3.1}Compare of the tri-boson process and the VBF process}

\begin{figure}[htbp]
\begin{center}
\includegraphics[width=0.58\hsize]{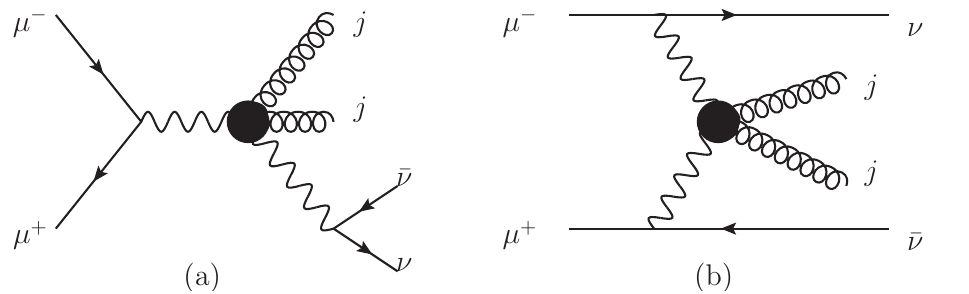}
\caption{\label{fig:fdnp}The Feynman diagrams of the gQGC contributions, the tri-boson contribution is shown in the left panel, while the VBF contribution is shown in the right panel.}
\end{center}
\end{figure}

The gQGCs can contribute to the process $\mu^+\mu^-\to jj\nu\bar{\nu}$ via both the VBF and tri-boson processes, the Feynman diagrams are shown in Fig.~\ref{fig:fdnp}.
When a new particle $X$ presents in the final states, the relative scaling between the VBF and annihilation contribution is \cite{AlAli:2021let}
\begin{equation}
\begin{split}
&\frac{\sigma ^{\rm NP}_{\rm VBF}}{\sigma ^{\rm NP}_{\rm ann}} \propto \alpha _W^2\frac{s}{m_X^2}\log^2\left(\frac{s}{M_V^2}\right)\log\left(\frac{s}{M_X^2}\right)
\end{split}
\label{eq.2.1}
\end{equation}
where $\sigma ^{\rm NP}_{\rm VBF}$ and $\sigma ^{\rm NP}_{\rm ann}$ are cross-sections of VBF and annihilation processes, respectively.
$M_{X,V}$ denotes the masses of the $X$ particle and the SM vector bosons, respectively.
At high energies, the VBF process is double-logarithmic enhanced.
However, it has been pointed out that, at lower energies the tri-boson process can be more sensitive to the dimension-8 operators \cite{triphoton}.
Therefore, it is necessary to compare the contribution of tri-boson and VBF for the case of gQGCs.

In the following, we consider the gQGC contribution to the process $\mu^+\mu^-\to \nu\bar{\nu}gg$.
For simplicity, defining $f_i\equiv 1/16 M_i^4$, the triboson and the VBF contributions~(denoting as $\sigma _{\rm tri}$ and $\sigma _{\rm VBF}$) at the leading order of $M_{Z,W}^2/s$ and at tree level are,
\begin{equation}
\begin{split}
&\sigma _{\rm tri}={\rm Br}(Z\to \nu\bar{\nu})\times \frac{e^2 s^3 }{8847360 \pi ^3 c_W^2 s_W^2}\\
&\times \left\{c_W^4 \left[24 f_0^2+4 f_3 (f_0+3 f_1+f_2)+8 f_0 f_1\right.\right.\\
&\left.\left.+12 f_0f_2+14 f_1^2+8f_1f_2+3 f_2^2+4 f_3^2\right]\right.\\
&\left.-2 c_W^2 s_W^2 \left[2 f_7 (f_0+3 f_1+f_2+2 f_3)+24 f_0 f_4+4 f_0 f_5\right.\right.\\
&\left.\left.+6 f_0 f_6+4 f_1 f_4+14 f_1 f_5+4 f_1 f_6\right.\right.\\
&\left.\left.+6 f_2 f_4+4 f_2 f_5+3 f_2f_6+2 f_3f_4+6 f_3f_5+2 f_3f_6\right]\right.\\
&\left.+5 s_W^4 \left[24 f_4^2+4 f_7 (f_4+3 f_5+f_6)+8 f_4 f_5+12 f_4f_6\right.\right.\\
&\left.\left.+14 f_5^2+8 f_5f_6+3 f_6^2+4 f_7^2\right]\right\}
\end{split}
\label{eq.2.2}
\end{equation}
and
\begin{equation}
\begin{split}
&\sigma _{\rm VBF}=\frac{e^4 s^3 }{4246732800000 \pi ^5 s_W^4}\\
&\times\left\{600 \log ^2\left(\frac{s}{16 M_W^2}\right) \left[800\left((6 f_0^2+f_0 (2 f_1+3 f_2+f_3)\right)\right.\right.\\
&\left.\left.+1492 f_1^2+52 f_1 (31 f_2+21 f_3)+603 f_2^2+806 f_2 f_3+473 f_3^2\right]\right.\\
&\left.-40 \log \left(\frac{s}{16 M_W^2}\right) \left[160800 f_0^2+26800 f_0 (2 f_1+3 f_2+f_3)\right.\right.\\
&\left.\left.+33944 f_1^2+4 f_1 (9491 f_2+5136 f_3)\right.\right.\\
&\left.\left.+16191 f_2^2+18982 f_2 f_3+11836 f_3^2\right]\right.\\
&\left.+4526400 f_0^2+754400 f_0 (2 f_1+3 f_2+f_3)+764788 f_1^2\right.\\
&\left.+877948 f_1 f_2+387588 f_1 f_3\right.\\
&\left.+408087 f_2^2+438974 f_2 f_3+285497 f_3^2\right\}
\end{split}
\label{eq.2.3}
\end{equation}
where $\sqrt{s}$ is the c.m. energies of the muon colliders, $s_W=\sin (\theta _W)$ and $c_W=\cos (\theta _W)$ with the weak mixing angle denoted as $\theta _W$, ${\rm Br}(Z\to \nu\bar{\nu}
)$ is taken as $20\%$.
In Eqs.~(\ref{eq.2.2}) and (\ref{eq.2.3}), the interference between $\sigma _{\rm tri}$ and $\sigma _{\rm VBF}$ are ignored.
The $\sigma _{\rm VBF}$ is obtained using the effective vector boson approximation~\cite{eva1,eva2,eva3},
\begin{equation}
\begin{split}
&\sigma_{\rm VBF} (\mu^+\mu^-\to \bar{\nu}\nu gg)\\
&=\sum _{\lambda _1\lambda _2\lambda _3\lambda _4}\int \rm{d}\xi _1\int \rm{d}\xi_2 f_{W_{\lambda _1}^-/\mu^-}(\xi _1)f_{W_{\lambda _2}^+/\mu^+}(\xi _2)\sigma _{W_{\lambda _1}^+W_{\lambda _2}^-\to gg}(\hat{s}),\\
&f_{W_{+1}^-/\mu _L^-}(\xi)=f_{W_{-1}^+/\mu_L^+}(\xi)=\frac{e^2}{8\pi^2s_W^2}\frac{(1-\xi)^2}{2\xi}\log \frac{\mu_f^2}{M_W},\\
&f_{W_{-1}^-/\mu _L^-}(\xi)=f_{W_{+1}^+/\mu_L^+}(\xi)=\frac{e^2}{8\pi^2s_W^2}\frac{1}{2\xi}\log \frac{\mu_f^2}{M_W},\\
&f_{W_{0}^-/\mu _L^-}(\xi)=f_{W_{0}^+/\mu_L^+}(\xi)=\frac{e^2}{8\pi^2s_W^2}\frac{1-\xi}{\xi},\\
&f_{W_{\lambda}^{\pm}/\mu _R^{\pm}}=0,\;\;\;\;f_{W_{\lambda}^{\pm}/\mu ^{\pm}}=\frac{f_{W_{\lambda}^{\pm}/\mu _L^{\pm}}+f_{W_{\lambda}^{\pm}/\mu _R^{\pm}}}{2},\\
\end{split}
\label{eq.2.4}
\end{equation}
where $\sqrt{\hat{s}}=\sqrt{\xi _1\xi _2 s}$ is the c.m. energy of $W^+W^-\to gg$, and  $\mu _f$ is the factorization scale set to be $\sqrt{\hat{s}}/4$~\cite{eva3}.

\begin{figure}[htbp]
\begin{center}
\includegraphics[width=0.58\hsize]{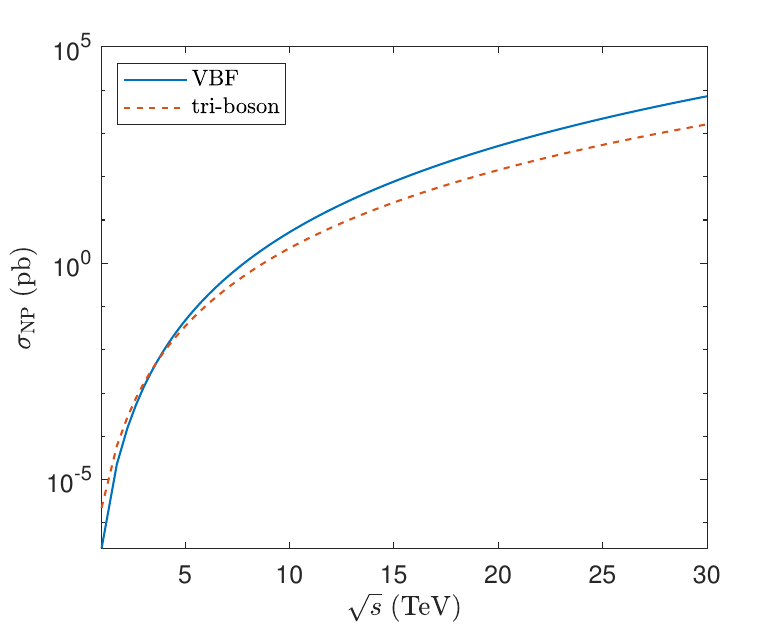}
\caption{\label{fig:cscompare}$\sigma _{\rm tri}$ in Eq.~(\ref{eq.2.2}) compared with $\sigma _{\rm VBF}$ in Eq.~(\ref{eq.2.3}). When $\sqrt{s}<5\;{\rm TeV}$, $\sigma _{\rm tri}$ is larger than $\sigma _{\rm VBF}$.}
\end{center}
\end{figure}
The numerical results of Eqs.~(\ref{eq.2.2}) and (\ref{eq.2.3}) are shown in Fig.~\ref{fig:cscompare}.
\textcolor[rgb]{0,0,1}{The detector simulation is not included until the Monte Carlo~(MC) simulation is applied.}
It can be seen that, the tri-boson contribution is larger than the VBF at about $\sqrt{s}<5\;{\rm TeV}$, at $\sqrt{s}=30\;{\rm TeV}$, $\sigma _{\rm VBF}$ is about $5$ times of $\sigma _{\rm tri}$. Therefore the contribution of tri-boson is not negligible.
In this paper, $\sigma _{\rm VBF}$, $\sigma _{\rm tri}$, and the intereference between them are considered as the signal of the gQGCs.

Apart from that, note that $O_{gT,4,5,6,7}$ operators does not contribute to the $W$-boson fusion processes.
Since the energies considered are mainly above $5\;{\rm TeV}$, in this paper, we only consider the contributions of the $O_{gT,0,1,2,3}$ operators.
However, we shall emphasis that the process $WW\to gg$ provides a chance to study the $O_{gT,0,1,2,3}$ operators separately, compared with the $pp\to \gamma\gamma$ and $pp\to Z\gamma$ processes where the contributions from $O_{gT,i}$ and $O_{gT,i+4}$ are propotional to each other~\cite{Ellis:2021dfa}.

\subsection{\label{sec3.2}Unitarity bound}

The SMEFT is not UV completed.
As an EFT, the SMEFT is only valid under a certain energy scale.
One of the signals when the SMEFT is no longer valid is the violation of unitarity.
With dimension-8 operators, the amplitude of the process $W^+W^-\to gg$ grows as $\hat{s}^2$ which leads to the violation of unitarity~\cite{unitarityHistory1,unitarityHistory2,unitarityHistory3} at large enough energy.
The partial wave unitarity bound~\cite{partialwaveunitaritybound,ubnew1,ubnew2} is often used to check whether the energy scale in considered is already invalid.
For the subprocess $W^+_{\lambda _1}W^-_{\lambda _2}\to g_{\lambda _3}g_{\lambda _4}$, where $\lambda _i$ correspond to the helicities, the amplitude can be expanded as~\cite{partialwaveexpansion}
\begin{equation}
\begin{split}
&\mathcal{M}(W^+_{\lambda _1}W^-_{\lambda _2}\to g_{\lambda _3}g_{\lambda _4})\\
&=8\pi \sum _{J}\left(2J+1\right)\sqrt{1+\delta _{\lambda _3,\lambda _4}}e^{\rm{i}(\lambda-\lambda ') \varphi}\rm{d}^J_{\lambda \lambda '}(\theta) T^J\\
\end{split}
\label{eq.3.1}
\end{equation}
where $\lambda = \lambda _1-\lambda _2$, $\lambda ' =\lambda _3-\lambda _4$, $\theta$ and $\phi$ are zenith and azimuth angles of the one of the gluons in the final state, respectively, and $d^J_{\lambda \lambda '}(\theta)$ are Wigner $d$-functions~\cite{partialwaveexpansion}.
The partial wave unitarity bound is $|T^J|\leq 2$~\cite{partialwaveunitaritybound} which is widely used in previous works~\cite{unitarity1,unitarity2,unitarity3,unitarity4,wastudy,zastudy,ntgc7,Yang:2022ilt,wwwwunitary,Yang:2022fhw}.

At leading order of $\hat{s}$, the relevant helicity amplitudes are
\begin{equation}
\begin{split}
&\mathcal{M}(W_+^+W_+^-\to g_+ g_+)=\frac{\hat{s}^2}{16} (4f_0+f_2+f_3),\\
&\mathcal{M}(W_+^+W_+^-\to g_- g_-)\\
&=\frac{\hat{s}^2}{64} \left(16 f_0+(2 f_1+f_3) \cos (2 \theta)+6 f_1+4 f_2-f_3\right),\\
&\mathcal{M}(W_+^+W_-^-\to g_+ g_-)=\frac{\hat{s}^2}{16} e^{2 \rm{i} \phi} \cos ^4\left(\frac{\theta}{2}\right) (2 f_1+f_2+f_3).\\
\end{split}
\label{eq.3.2}
\end{equation}
There are other helicity amplitudes which lead to same unitarity bounds as those derived from Eq.~(\ref{eq.3.2}), and therefore are not presented for simplicity.
Assuming one operator at a time, the tightest bounds are,
\begin{equation}
\begin{split}
&\frac{\hat{s}^2 |f_0|}{32\sqrt{2}\pi}<2,\;\;\;
\frac{\hat{s}^2 |f_1|}{96\sqrt{2}\pi}<2,\;\;\;
\frac{\hat{s}^2 |f_{2,3}|}{128\sqrt{2}\pi}<2.
\end{split}
\label{eq.3.3}
\end{equation}

\begin{table}[htbp]
\centering
\begin{tabular}{c|c|c|c|c}
\hline
 $\sqrt{\hat{s}}\;({\rm TeV})$  & 3  & 10 & 14 & 30 \\
\hline
 $|f_0|\;({\rm TeV}^{-4})$ &  3.5 &  0.028 & 0.0074 & 0.00035 \\
\hline
 $M_0\;({\rm GeV})$ &  365.56 &  1222.31 & 1704.76 & 3655.55 \\
\hline
 $|f_1|\;({\rm TeV}^{-4})$ &  10.5 &  0.085 & 0.022 &  0.001 \\
\hline
 $M_1\;({\rm GeV})$ &  277.76 &  926.01 & 1298.27 & 2811.71 \\
\hline
 $|f_{2,3}|\;({\rm TeV}^{-4})$ & 14.0 & 0.114 & 0.030 &  0.004 \\
\hline
 $M_{2,3}\;({\rm GeV})$ &  258.49 &  860.49 & 1201.41 & 1988.18 \\
\hline
\end{tabular}
\caption{The upper bounds of coefficients $|f_i|$ and lower bounds of $M_i$ in the sense of partial wave unitarity at different energies.}
\label{table:unitarity}
\end{table}

For the subprocess $W^+W^-\to gg$, $\hat{s}\leq s$, therefore we consider the largest possible $\hat{s}$, the unitarity bounds on the coefficients at different energies are listed in Table~\ref{table:unitarity}.
The energies are chosen as the muon colliders \cite{Black:2022cth,Accettura:2023ked}.

\subsection{\label{sec3.3}Event selection strategy}

\begin{figure}[htbp]
\begin{center}
\includegraphics[width=0.58\hsize]{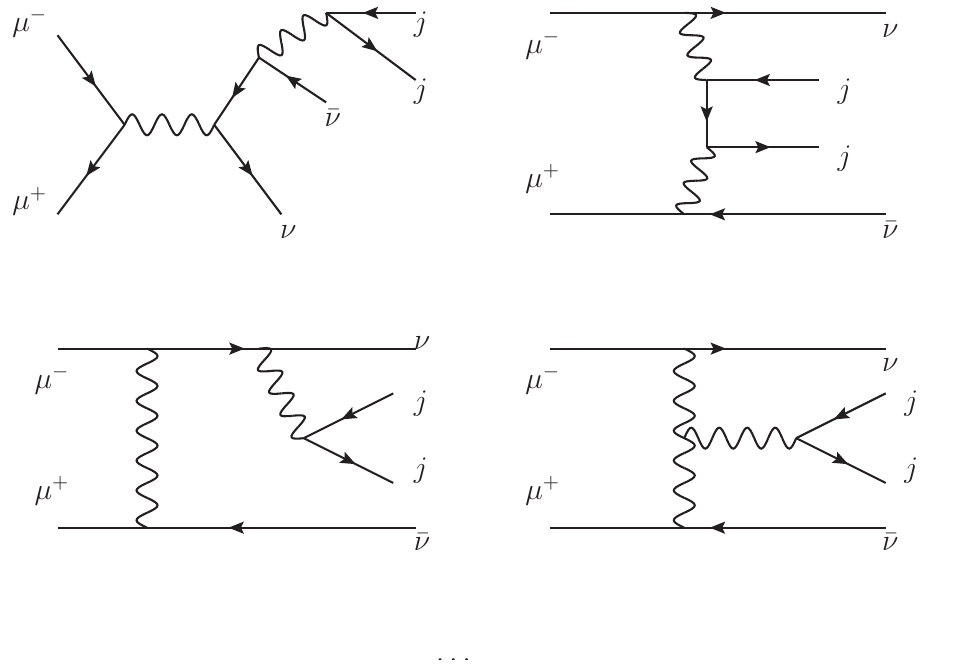}
\caption{\label{fig:fdsm}Typical Feynman diagrams of the SM contribution.}
\end{center}
\end{figure}

The background in this study is the SM contribution to $\mu^+\mu^-\to jj \nu\bar{\nu}$.
The process $\mu^+\mu^-\to j j$ is suppressed by the s-channel propagator, and can be further suppressed by the cuts on the missing energy, therefore is ignored.
The process $\mu^+\mu^-\to j j\nu\nu\bar{\nu}\bar{\nu}$ is suppressed by more EW couplings, therefore is also ignored.

The typical Feynman diagrams for the background at tree level are shown in Fig.~\ref{fig:fdsm}.
An important feature is that the two jets are both from quarks, which is different from the case of gQGCs where the two jets are both from the gluons.
As a consequence, there is no interference term between the SM and the gQGCs.
Once the efficiencies of the event selection strategy for both the background and the signal are obtained, the cross-section can be read out.

To investigate the event selection strategy, the MC simulation is carried out with the help of \verb"MadGraph5_aMC@NLO"~\cite{madgraph,feynrules} toolkit including a parton shower using \verb"Pythia82"~\cite{pythia}.
The standard cuts are used as the default.
The parton distribution function is NNPDF2.3~\cite{NNPDF}.
A fast detector simulation is then applied using \verb"Delphes"~\cite{delphes} with the muon collider card. 
The analyses of the signal and the background are archived by using \verb"MLAnalysis"~\cite{Guo:2023nfu}. 
\textcolor[rgb]{0,0,1}{In the event generation, the complete syntax $\mu^+\mu^-\to jj\nu_l\bar{\nu}_l$ is used, and the standard cuts are set as the default of ``MadGraph5\_aMC@NLO'', the relevant cuts for jets are transverse momentum $p_T^j > 20$ GeV and pseudo-rapidity $\eta_j < 5.0$, and $\Delta R_{jj}>0.4$ where $\Delta R=\sqrt{\Delta \phi _{jj}^2+\Delta \eta_{jj}^2}$ where $\Delta \phi _{jj}$ and $\Delta \eta_{jj}$ are difference between the azimuth angles and pseudo-rapidities of any two jets.}

\begin{table}[htbp]
\centering
\begin{tabular}{c|c|c|c|c}
\hline
 $\sqrt{s}\;({\rm TeV})$  & 3 & 10 & 14 & 30 \\
\hline
 $\sigma _{\rm SM}\;({\rm pb})$ & $0.8688$ &  $1.4548$ &  $1.6087$ &  $1.8988$ \\
\hline
 $|f_0|\;({\rm TeV}^{-4})$ & 1 & 0.012 &  0.004 &  0.00035 \\
 $\sigma _{gT,0}\;({\rm pb})$ & $0.00321$ &  $0.00115$ &  $0.00112$ &  $0.00114$ \\
\hline
 $|f_1|\;({\rm TeV}^{-4})$ & 1.5 & 0.02 &  0.007 &  0.0006 \\
 $\sigma _{gT,1}\;({\rm pb})$ & $0.00355$ &  $0.00138$ &  $0.00144$ &  $0.00134$ \\
\hline
 $|f_{2,3}|\;({\rm TeV}^{-4})$ & 3 & 0.03 &  0.012 &  0.0012 \\
 $\sigma _{gT,2}\;({\rm pb})$ & $0.00383$ &  $0.000956$ &  $0.00134$ &  $0.00178$ \\
 $\sigma _{gT,3}\;({\rm pb})$ & $0.00412$ &  $0.000924$ &  $0.00127$ &  $0.00161$ \\
\hline
\end{tabular}
\caption{The cross-sections of the SM contribution and the upper bounds of coefficients $|f_i|$ used in the phenomenological study. The cross-sections of the gQGCs are also shown.}
\label{table:coefficientscan}
\end{table}

At the energies of the muon colliders, the SM cross-sections~(denoted as $\sigma _{\rm SM}$) are obtained and listed in Table~\ref{table:coefficientscan}.
In the SM, the cross-section grows slowly with the energy.

The signal events are generated by assuming one operator at a time.
The sensitivity of the process $\mu^+\mu^-\to j j \nu\bar{\nu}$ to the gQGCs can be estimated with respect to the significance defined as~\cite{Cowan:2010js,pdgss:2020}
\begin{equation}
\begin{split}
&\mathcal{S}_{stat}=\sqrt{2 \left[(N_{\rm bg}+N_{s}) \ln (1+N_{s}/N_{\rm bg})-N_{s}\right]}\;,
\end{split}
\label{eq.3.4}
\end{equation}
where $N_s=N_{\rm aQGC}-N_{\rm SM}$ and $N_{\rm bg}=N_{\rm SM}$.
The $N_{\rm bg}$ can be obtained by $\sigma _{\rm SM}$ and the luminosities of the muon colliders, where the $N_s$ can be provisionally estimated using $\sigma _{\rm tri} + \sigma _{\rm VBF}$ where $\sigma _{\rm tri}$ and $\sigma _{\rm VBF}$ are given in Eqs.~(\ref{eq.2.2}) and (\ref{eq.2.3}).
The coefficients are chosen so that the signal significance of the VBF contributions are about $S_{\rm stat}=2\sim 3$ before cuts at luminosity $L=1\;{\rm ab}^{-1}$, $L=10\;{\rm ab}^{-1}$ for $\sqrt{s}=3\;{\rm TeV}$ and $\sqrt{s}\geq 10\;{\rm TeV}$, respectively~\cite{Black:2022cth,Accettura:2023ked}.
The coefficients at different energies are listed in Table~\ref{table:coefficientscan}.
Note that they are also chosen to be within the partial wave unitarity bounds in Table~\ref{table:unitarity}.
The contributions of the gQGCs~(denoted as $\sigma _{gT,i}$) obtained by MC are also shown in Table~\ref{table:coefficientscan}.

\begin{figure}[htbp]
\begin{center}
\includegraphics[width=0.48\hsize]{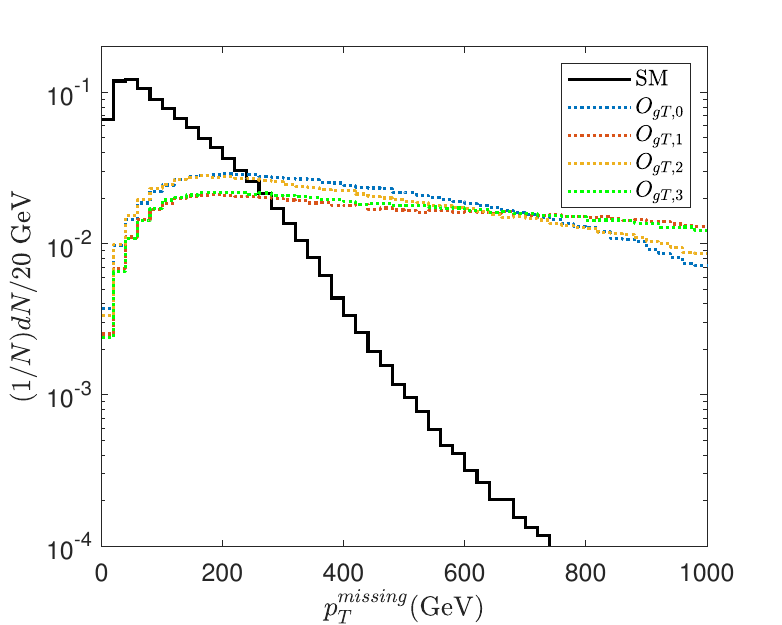}
\includegraphics[width=0.48\hsize]{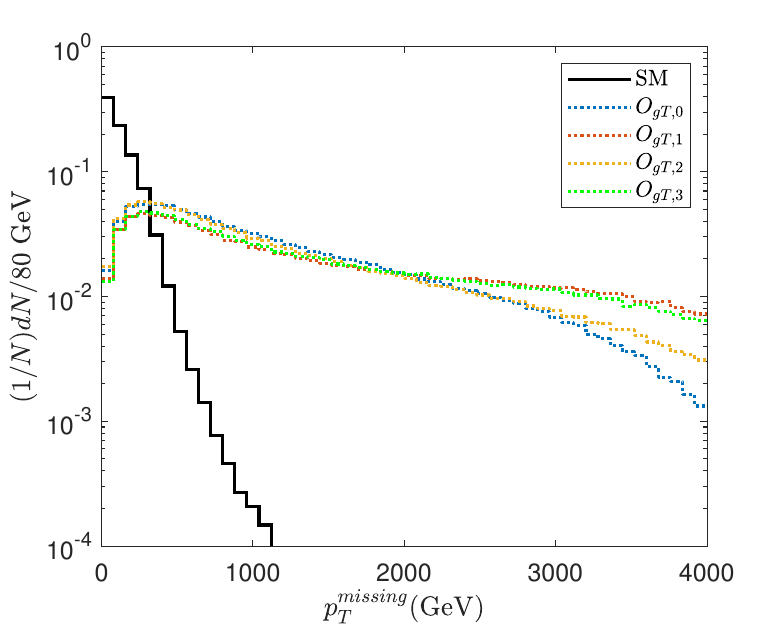}\\
\includegraphics[width=0.48\hsize]{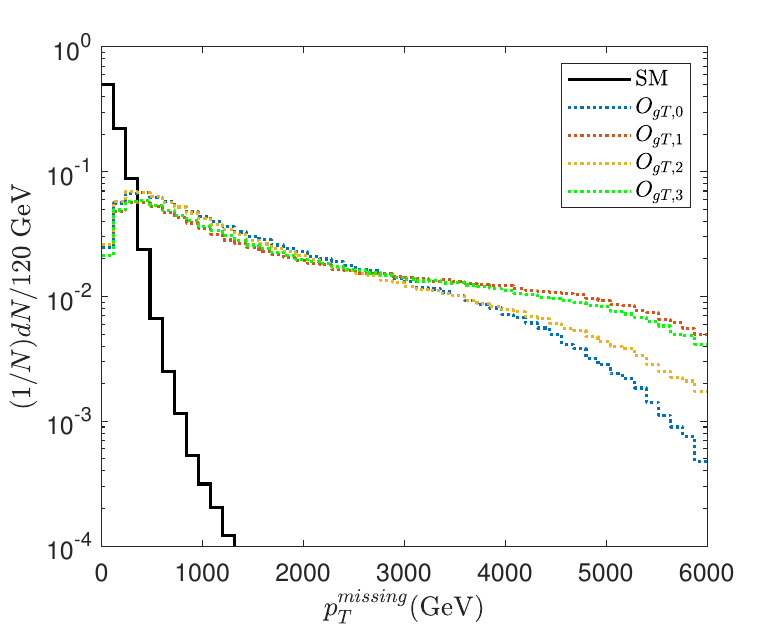}
\includegraphics[width=0.48\hsize]{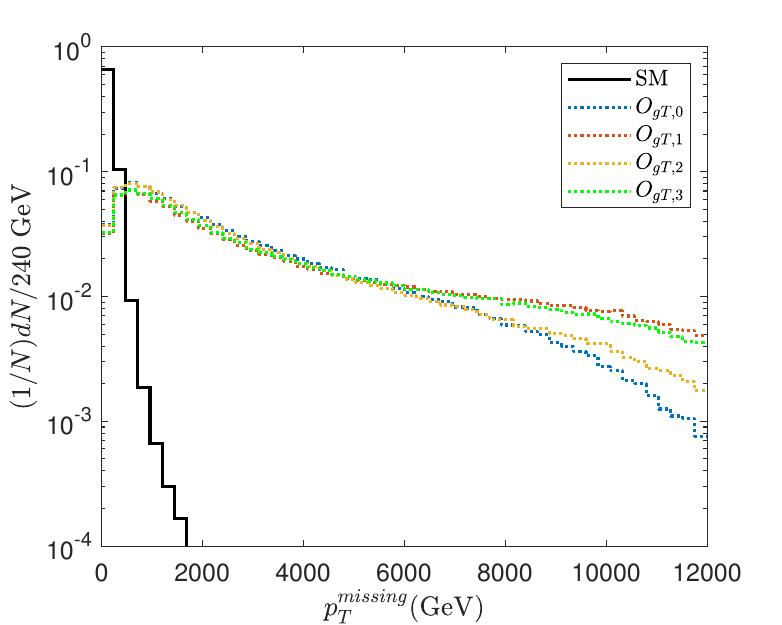}
\caption{\label{fig:pt}The normalized distributions of $\slashed{p}_T$ for the SM and gQGCs at different energies.
The top left panel corresponds to $\sqrt{s}=3\;{\rm TeV}$, the top right panel corresponds to $\sqrt{s}=10\;{\rm TeV}$, the bottom left panel corresponds to $\sqrt{s}=14\;{\rm TeV}$, and the bottom right panel corresponds to $\sqrt{s}=30\;{\rm TeV}$.}
\end{center}
\end{figure}
To investigate the kinematic features, we require that the final states contain at least two jets, which is denoted as the $N_j$ cut. 
To suppress the events from the process $\mu^+\mu^-\to j j$, we also require a minimal $\slashed{p}_T$ for the events, where $\slashed{p}_T$ is the transverse missing energy.
$\slashed{p}_T$ can also be used to suppress the events from the SM contribution to $\mu^+\mu^-\to j j\nu \bar{\nu}$.
This is because the signal grows with the energy, so that for the tri-boson signal events, the neutrinos are typically from a highly boosted $Z$ boson.
Since neutrinos play a role similar to that of the residual jets of the VBF processes at a hadron collider, similar as the standard VBF cut~\cite{Rauch:2016pai}, the neutrinos are expected to be back-to-back.
For the VBF signal events, although the neutrinos are back-to-back, the net transpose momentum is still typically larger than the case of the SM.
The normalized distributions of $\slashed{p}_T$ for the backgrounds and signals are shown in Fig.~\ref{fig:pt}.
We require the events to have a large $\slashed{p}_T$, which is denoted as the $\slashed{p}_T$ cut.

\begin{figure}[htbp]
\begin{center}
\includegraphics[width=0.48\hsize]{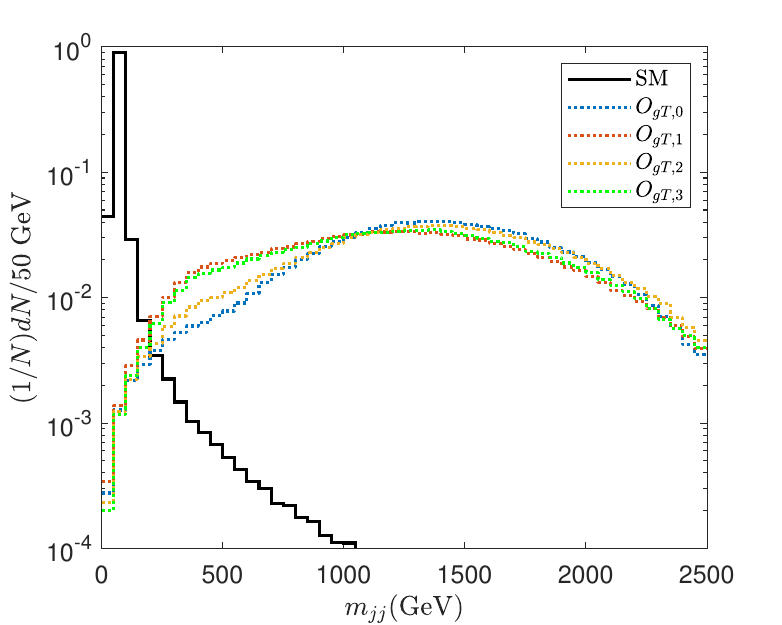}
\includegraphics[width=0.48\hsize]{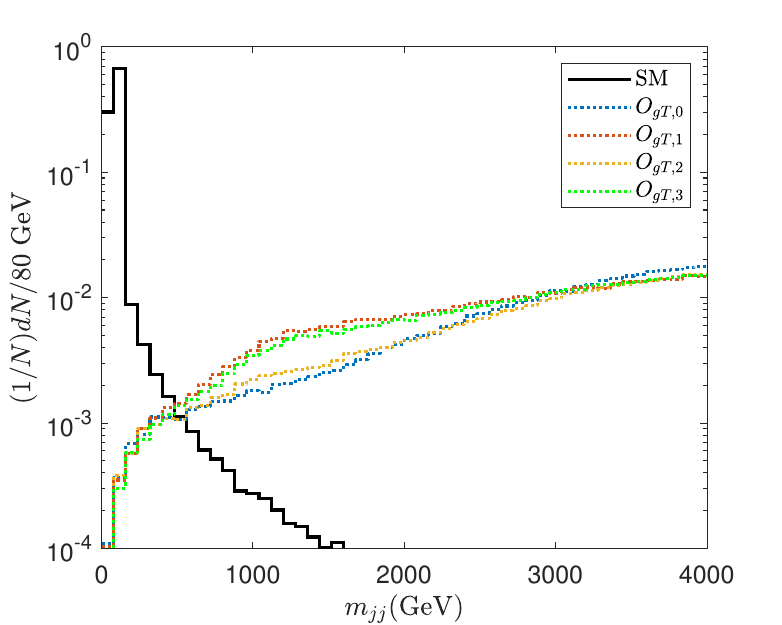}\\
\includegraphics[width=0.48\hsize]{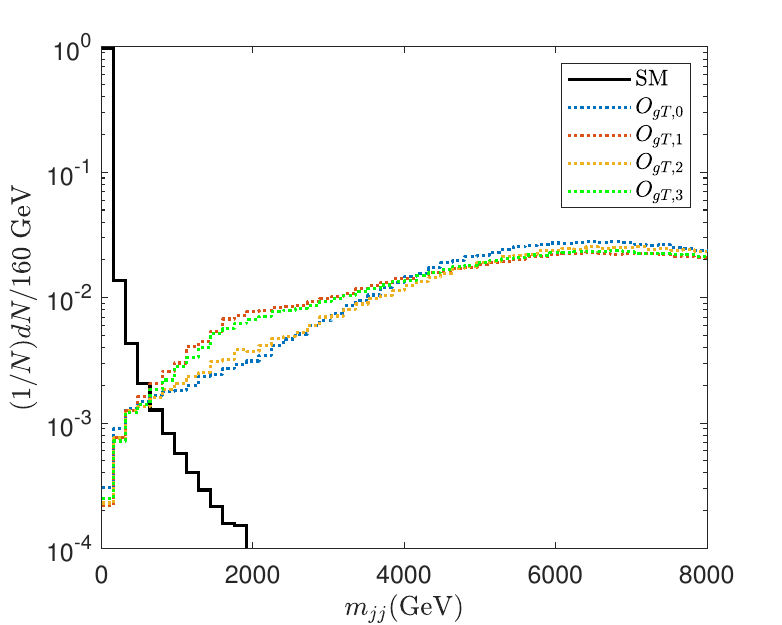}
\includegraphics[width=0.48\hsize]{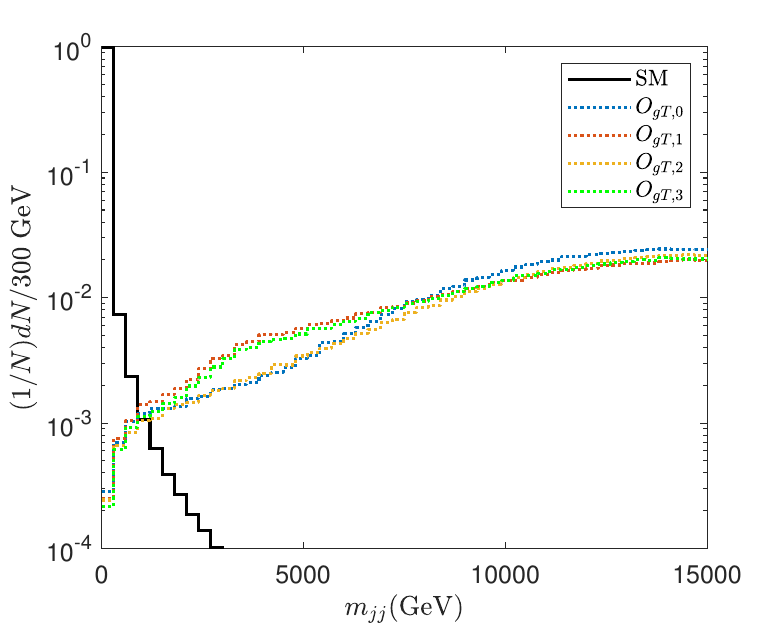}
\caption{\label{fig:mjj}Same as Fig.~\ref{fig:pt} but for $m_{jj}$.}
\end{center}
\end{figure}
Due to the same reason, for both the tri-boson and VBF signal events, the jets are energetic.
As a result, the invariant mass of the hardest two jets~(denoted as $m_{jj}$) should be large for the signal events.
The normalized distributions of $m_{jj}$ for the background and signal are shown in Fig.~\ref{fig:mjj}.
We require the events to have a large $m_{jj}$, which is denoted as the $m_{jj}$ cut.

\begin{table}[htbp]
\centering
\begin{tabular}{c|c|c}
\hline
 $\sqrt{s}$ & $\slashed{p}_T$ & $m_{jj}$  \\
 $({\rm TeV})$ &  &  \\
\hline
 $3$   &  $>50\;{\rm GeV}$ & $>1\;{\rm TeV}$ \\
 $10$  &  $>100\;{\rm GeV}$ & $>3\;{\rm TeV}$ \\
 $14$  &  $>100\;{\rm GeV}$ & $>5\;{\rm TeV}$ \\
 $30$  &  $>200\;{\rm GeV}$ & $>10\;{\rm TeV}$ \\
\hline
\end{tabular}
\caption{The event selection strategies at different energies.}
\label{table:cuts}
\end{table}

\begin{table}[htbp]
\centering
\begin{tabular}{c|c|c|c|c|c|c}
\hline
 $\sqrt{s}$ & cut & SM & $O_{gT,0}$ & $O_{gT,1}$ & $O_{gT,2}$ & $O_{gT,3}$ \\
 $({\rm TeV})$ & & (fb) & (fb) & (fb) & (fb) & (fb) \\
\hline
      & $N_j$          & $722.9$  & $3.21$  & $3.52$  & $3.82$  & $4.10$  \\
 $3$  & $\slashed{p}_T$ & $536.7$  & $3.15$  & $3.47$  & $3.74$  & $4.04$  \\
      & $m_{jj}$        & $0.885$  & $2.49$  & $2.27$  & $2.85$  & $2.74$  \\
      & $\epsilon$      & $0.102\%$ & $77.6\%$ & $63.9\%$ & $74.4\%$ & $66.5\%$ \\
\hline
 $\sqrt{s}$ & cut & SM & $O_{gT,0}$ & $O_{gT,1}$ & $O_{gT,2}$ & $O_{gT,3}$ \\
 $({\rm TeV})$ & & (fb) & (fb) & (fb) & (fb) & (fb) \\
\hline
      & $N_j$           & $1191.9$ & $1.15$  & $1.37$  & $0.955$ & $0.921$ \\
 $10$ & $\slashed{p}_T$ & $508.2$  & $1.12$  & $1.34$  & $0.930$ & $0.902$ \\
      & $m_{jj}$        & $0.256$  & $0.961$ & $1.06$  & $0.799$ & $0.728$ \\
      & $\epsilon$      & $0.0176\%$ & $83.6\%$ & $76.8\%$ & $83.6\%$ & $78.8\%$ \\
\hline
 $\sqrt{s}$ & cut & SM & $O_{gT,0}$ & $O_{gT,1}$ & $O_{gT,2}$ & $O_{gT,3}$ \\
 $({\rm TeV})$ & & (fb) & (fb) & (fb) & (fb) & (fb) \\
\hline
      & $N_j$           & $1315.5$ & $1.12$  & $1.43$  & $1.34$  & $1.26$  \\
 $14$ & $\slashed{p}_T$ & $531.9$  & $1.10$  & $1.42$  & $1.31$  & $1.24$  \\
      & $m_{jj}$        & $0.105$  & $0.846$ & $1.02$  & $1.03$  & $0.916$ \\
      & $\epsilon$      & $0.00653\%$ & $75.5\%$ & $70.8\%$ & $76.9\%$ & $72.1\%$ \\
\hline
 $\sqrt{s}$ & cut & SM & $O_{gT,0}$ & $O_{gT,1}$ & $O_{gT,2}$ & $O_{gT,3}$ \\
 $({\rm TeV})$ & & (fb) & (fb) & (fb) & (fb) & (fb) \\
\hline
      & $N_j$           & $1542.4$ & $1.14$  & $1.33$  & $1.78$  & $1.61$  \\
 $30$ & $\slashed{p}_T$ & $256.4$  & $1.11$  & $1.30$  & $1.73$  & $1.57$  \\
      & $m_{jj}$        & $0.0532$ & $0.914$ & $1.04$  & $1.46$  & $1.27$  \\
      & $\epsilon$      & $0.00280\%$ & $80.2\%$ & $77.6\%$ & $82.0\%$ & $78.9\%$ \\
\end{tabular}
\caption{The contributions of the SM and gQGCs after cuts. The efficiency of the cuts are shown in the last row.}
\label{table:cutflow}
\end{table}

At different energies, we use different cuts.
The event selection strategies are summarized in Table~\ref{table:cuts}.
The cross-sections after cuts are listed in Table~\ref{table:cutflow}.
The efficiency of the cuts~(denoted as $\epsilon$) are also shown in Table~\ref{table:cutflow}.
It can be seen that, the event selection strategies can suppressed the background significantly.

\section{\label{sec4}Constraints on the coefficients}

\begin{table*}
\begin{center}
\begin{tabular}{c|c|c|c|c|c}
\hline
    & $\mathcal{S}_{stat}$ & $3$ TeV & $10$ TeV & $14$ TeV & $30$ TeV \\
    & & $1\;{\rm ab}^{-1}$ & $10\;{\rm ab}^{-1}$ & $10\;{\rm ab}^{-1}$ & $10\;{\rm ab}^{-1}$ \\
\hline
                             & 2 & $<155$ & $<1.24$ & $<0.351$ & $<0.025$  \\
$|f_{0}|$
                             & 3 & $<191$ & $<1.52$ & $<0.432$ & $<0.03$ \\
$(10^{-3}{\rm TeV}^{-4})$
                             & 5 & $<248$ & $<1.96$ & $<0.561$ & $<0.04$ \\
\hline
                             & 2 & $>0.796$ & $>2.67$ & $>3.65$  & $>7.07$ \\
$M_0$
                             & 3 & $>0.756$ & $>2.53$ & $>3.47$  & $>6.71$ \\
$({\rm TeV})$
                             & 5 & $>0.709$ & $>2.38$ & $>3.25$  & $>6.29$ \\
\hline
                             & 2 & $<244$ & $<1.96$ & $<0.561$ & $<0.040$  \\
$|f_{1}|$
                             & 3 & $<300$ & $<2.41$ & $<0.689$ & $<0.049$ \\
$(10^{-3}{\rm TeV}^{-4})$
                             & 5 & $<389$ & $<3.11$ & $<0.893$ & $<0.064$ \\
\hline
                             & 2 & $>0.711$ & $>2.38$ & $>3.25$  & $>6.28$ \\
$M_1$
                             & 3 & $>0.675$ & $>2.26$ & $>3.09$  & $>5.96$ \\
$({\rm TeV})$
                             & 5 & $>0.633$ & $>2.12$ & $>2.89$  & $>5.58$ \\
\hline
                             & 2 & $<436$ & $<3.39$ & $<0.957$ & $<0.068$  \\
$|f_{2}|$
                             & 3 & $<535$ & $<4.16$ & $<1.17$ & $<0.083$ \\
$(10^{-3}{\rm TeV}^{-4})$
                             & 5 & $<695$ & $<5.38$ & $<1.52$ & $<0.109$ \\
\hline
                             & 2 & $>0.615$ & $>2.07$ & $>2.84$  & $>5.07$ \\
$M_2$
                             & 3 & $>0.585$ & $>1.97$ & $>2.70$  & $>5.23$ \\
$({\rm TeV})$
                             & 5 & $>0.548$ & $>1.85$ & $>2.53$  & $>4.90$ \\
\hline
                             & 2 & $<445$ & $<3.55$ & $<1.01$ & $<0.073$  \\
$|f_{3}|$
                             & 3 & $<546$ & $<4.35$ & $<1.25$ & $<0.090$ \\
$(10^{-3}{\rm TeV}^{-4})$
                             & 5 & $<709$ & $<5.64$ & $<1.62$ & $<0.116$ \\
\hline
                             & 2 & $>0.612$ & $>2.05$ & $>2.80$  & $>5.41$ \\
$M_3$
                             & 3 & $>0.582$ & $>1.95$ & $>2.66$  & $>5.14$ \\
$({\rm TeV})$
                             & 5 & $>0.545$ & $>1.82$ & $>2.49$  & $>4.81$ \\
\hline
\end{tabular}
\end{center}
\caption{\label{tab.constraint1}The projected sensitivities on the aQGC coefficients at the muon colliders with different c.m. energies and integrated luminosities for ``conservative'' case.
}
\end{table*}

\begin{table}
\begin{center}
\begin{tabular}{c|c|c|c}
\hline
    & $\mathcal{S}_{stat}$ & $14$ TeV & $30$ TeV \\
    & & $20\;{\rm ab}^{-1}$ & $90\;{\rm ab}^{-1}$ \\
\hline
                             & 2 & $<2.95$ & $<0.144$ \\
$|f_{0}|$
                             & 3 & $<3.63$ & $<0.176$ \\
$(10^{-4}{\rm TeV}^{-4})$
                             & 5 & $<4.70$ & $<0.228$ \\
\hline
                             & 2 & $>3.81$ & $>8.12$ \\
$M_0$
                             & 3 & $>3.62$ & $>7.71$ \\
$({\rm TeV})$
                             & 5 & $>3.40$ & $>7.23$ \\
\hline
                             & 2 & $<4.71$ & $<0.231$ \\
$|f_{1}|$
                             & 3 & $<5.78$ & $<0.284$ \\
$(10^{-4}{\rm TeV}^{-4})$
                             & 5 & $<7.49$ & $<0.367$ \\
\hline
                             & 2 & $>3.39$ & $>7.21$ \\
$M_1$
                             & 3 & $>3.23$ & $>6.85$ \\
$({\rm TeV})$
                             & 5 & $>3.02$ & $>6.42$ \\
\hline
                             & 2 & $<8.03$ & $<0.390$ \\
$|f_{2}|$
                             & 3 & $<9,86$ & $<0.479$ \\
$(10^{-4}{\rm TeV}^{-4})$
                             & 5 & $<12.8$ & $<0.619$ \\
\hline
                             & 2 & $>2.97$ & $>6.33$ \\
$M_2$
                             & 3 & $>2.82$ & $>6.01$ \\
$({\rm TeV})$
                             & 5 & $>2.65$ & $>5.64$ \\
\hline
                             & 2 & $<8.52$ & $<0.419$ \\
$|f_{3}|$
                             & 3 & $<10.5$ & $<0.513$ \\
$(10^{-4}{\rm TeV}^{-4})$
                             & 5 & $<13.5$ & $<0.664$ \\
\hline
                             & 2 & $>2.93$ & $>6.22$ \\
$M_3$
                             & 3 & $>2.78$ & $>5.91$ \\
$({\rm TeV})$
                             & 5 & $>2.61$ & $>5.54$ \\
\hline
\end{tabular}
\end{center}
\caption{\label{tab.constraint2}The projected sensitivities on the aQGC coefficients at the muon colliders with different c.m. energies and integrated luminosities for ``optimistic'' case.
}
\end{table}

Assuming one operator at a time, without the interference between the SM and the gQGCs, the cross-section after cuts can be expressed as,
\begin{equation}
\begin{split}
&\sigma = \epsilon _{\rm SM}\sigma _{\rm SM} + \epsilon _{gT,i}\frac{f_i^2}{\tilde{f}_i^2}\sigma _{gT,i}(\tilde{f}_i)
\label{eq.4.1}
\end{split}
\end{equation}
where $\sigma _{\rm SM}$ and $\sigma _{gT,i}(\tilde{f}_i)$ are the cross-sections of the SM and the gQGCs at $\tilde{f}_i$, and $\epsilon_{\rm SM}$ and $\epsilon _{gT,i}$ are the cut efficiencies of the SM and gQGC events, respectively.
Taking $\tilde{f}_i$ as the ones in Table~\ref{table:coefficientscan}, numerical results of $\sigma _{\rm SM}$ and $\sigma _{gT,i}$ are listed in Table~\ref{table:coefficientscan}, and $\epsilon_{\rm SM}$ and $\epsilon _{gT,i}$ are listed in Table~\ref{table:cutflow}.
The expected constraints at muon colliders can be estimated by the signal significance defined in Eq.~(\ref{eq.3.4}).
The integrated luminosities for both the ``conservative'' and ``optimistic'' cases are considered \cite{Black:2022cth,Accettura:2023ked}.
As a result, we can obtain the projected sensitivities on $f_{i}$ and $M_i$ by taking $2\sigma$, $3\sigma$ or $5\sigma$ significance.
The results are shown in Tables~\ref{tab.constraint1} and \ref{tab.constraint2}.
The expected constraints on $f_{2}$ and $f_3$ are close to each other, which is indicated by the fact that the leading helicity amplitudes for $O_{2,3}$ are the same.

Compared with the expected constraints from the LHC in Eq.~(\ref{eq.1.1}), the muon collider at $\sqrt{s}\geq 10\;{\rm TeV}$ have tighter constraints.
Since the operators contributing to the gQGCs are dimension-8, not surprisingly, higher energy muon colliders are more advantageous in detecting the gQGCs.
Taking $O_{gT,0}$ as an example, at $\sqrt{s}=3\;{\rm TeV}$, the sensitivity will exceed that of the LHC when the luminosity is about $70\;{\rm ab}^{-1}$.
However, for the luminosities in the ``optimistic'' case, i.e., $L=1,10,20$ and $90\;{\rm ab}^{-1}$ for $\sqrt{s}=3,10,14$ and $30\;{\rm TeV}$, respectively, the $2\sigma$ lower bounds on $M_0$ is about $27\%$ of the c.m. energies of the muon collider, which indicates that the constraint on $M_0$ will exceed that of the LHC at a $\sqrt{s}=4\;{\rm TeV}$ muon collider.
Meanwhile, in the same case, the $2\sigma$ lower bounds on $M_{1,2,3}$ are about $24\%$, $21\%$ and $20\sim 21\%$ of $\sqrt{s}$, which are consistent with Eq.~(\ref{eq.1.1}) in order of sensitivity.
The above result can also be compared with the case of futrue hadron colliders such as the HL-LHC.
In Ref.~\cite{Ellis:2021dfa}, taking $O_{gT,0}$ as an example, the $2\sigma$ constraint on $M_0$ by the $pp\to \gamma\gamma$ channel is about $15\%$ of $\sqrt{s}$, and the combined~(combined of $pp\to \gamma\gamma$, $pp\to \ell^+\ell^-\gamma$, $pp\to \nu\bar{\nu}\gamma$ and $pp\to q\bar{q}\gamma$ channels) constraint on $M_0$ is about $22\%$ of $\sqrt{s}$.
Thus, the sensitivities of the muon colliders to the gQGCs are competitive to the future hardron colliders, and are even better at the same c.m. energies.

\section{\label{sec5}Summary}

The muon colliders which are also called vector boson colliders are idea places to study the quartic gauge couplings.
In this paper, we investigate the capability of the future muon colliders to detect the gQGC in the framework of SMEFT.
The cross-sections of the process $\mu^+\mu^-\to j j \nu\bar{\nu}$ with gQGCs present are calculated.

The gQGCs can affect the process $\mu^+\mu^-\to j j \nu\bar{\nu}$ via both tri-boson and VBF contributions.
The tri-boson contribution is larger than the VBF when $\sqrt{s}< 5\;{\rm TeV}$, and is about $1/5$ of the VBF at $\sqrt{s}=30\;{\rm TeV}$.
Besides, the VBF process can be affected by $O_{gT,0,1,2,3}$.
Therefore, we focus on $O_{gT,0,1,2,3}$ in this paper.

With partial wave unitarity bounds considered, the event selection strategy as well as the expected constraints on the operator coefficients are studied using MC simulation.
For the luminosities in the ``optimistic'' case,
$2\sigma$ lower bounds on $M_{0,1,2,3}$ are about $27\%$, $24\%$, $21\%$ and $20\sim 21\%$ of the c.m. energies of the muon colliders, respectively.
Our result indicate that the muon colliders with $\sqrt{s}>4\;{\rm TeV}$ can be more sensitive to the gQGCs than the LHC.

\addcontentsline{toc}{chapter}{Acknowledgment}
\section*{Acknowledgment}

\noindent
This work was supported in part by the National Natural Science Foundation of China under Grants Nos. 11905093 and 12147214, the Natural Science Foundation of the Liaoning Scientific Committee No.~LJKZ0978.

\bibliography{WWgg}

\providecommand{\href}[2]{#2}\begingroup\raggedright\begin{thebibliography}{10}

\bibitem{johnellis}
J.~Ellis, \emph{{Outstanding questions: Physics beyond the Standard Model}},
  \href{https://doi.org/10.1098/rsta.2011.0452}{\emph{Phil. Trans. Roy. Soc.
  Lond. A} {\bfseries 370} (2012) 818}.

\bibitem{weinberg}
S.~Weinberg, \emph{{Baryon and Lepton Nonconserving Processes}},
  \href{https://doi.org/10.1103/PhysRevLett.43.1566}{\emph{Phys. Rev. Lett.}
  {\bfseries 43} (1979) 1566}.

\bibitem{SMEFTReview1}
B.~Grzadkowski, M.~Iskrzynski, M.~Misiak and J.~Rosiek, \emph{{Dimension-Six
  Terms in the Standard Model Lagrangian}},
  \href{https://doi.org/10.1007/JHEP10(2010)085}{\emph{JHEP} {\bfseries 10}
  (2010) 085} [\href{https://arxiv.org/abs/1008.4884}{{\ttfamily 1008.4884}}].

\bibitem{SMEFTReview2}
S.~Willenbrock and C.~Zhang, \emph{{Effective Field Theory Beyond the Standard
  Model}}, \href{https://doi.org/10.1146/annurev-nucl-102313-025623}{\emph{Ann.
  Rev. Nucl. Part. Sci.} {\bfseries 64} (2014) 83}
  [\href{https://arxiv.org/abs/1401.0470}{{\ttfamily 1401.0470}}].

\bibitem{SMEFTReview3}
E.~Masso, \emph{{An Effective Guide to Beyond the Standard Model Physics}},
  \href{https://doi.org/10.1007/JHEP10(2014)128}{\emph{JHEP} {\bfseries 10}
  (2014) 128} [\href{https://arxiv.org/abs/1406.6376}{{\ttfamily 1406.6376}}].

\bibitem{BI}
M.~Born and L.~Infeld, \emph{{Foundations of the new field theory}},
  \href{https://doi.org/10.1098/rspa.1934.0059}{\emph{Proc. Roy. Soc. Lond. A}
  {\bfseries 144} (1934) 425}.

\bibitem{Ellis:2018cos}
J.~Ellis and S.-F. Ge, \emph{{Constraining Gluonic Quartic Gauge Coupling
  Operators with gg\textrightarrow{}\ensuremath{\gamma}\ensuremath{\gamma}}},
  \href{https://doi.org/10.1103/PhysRevLett.121.041801}{\emph{Phys. Rev. Lett.}
  {\bfseries 121} (2018) 041801}
  [\href{https://arxiv.org/abs/1802.02416}{{\ttfamily 1802.02416}}].

\bibitem{d81}
B.~Henning, X.~Lu, T.~Melia and H.~Murayama, \emph{{2, 84, 30, 993, 560, 15456,
  11962, 261485, ...: Higher dimension operators in the SM EFT}},
  \href{https://doi.org/10.1007/JHEP08(2017)016}{\emph{JHEP} {\bfseries 08}
  (2017) 016} [\href{https://arxiv.org/abs/1512.03433}{{\ttfamily
  1512.03433}}].

\bibitem{looportree}
C.~Arzt, M.~Einhorn and J.~Wudka, \emph{{Patterns of deviation from the
  standard model}},
  \href{https://doi.org/10.1016/0550-3213(94)00336-D}{\emph{Nucl. Phys. B}
  {\bfseries 433} (1995) 41}
  [\href{https://arxiv.org/abs/hep-ph/9405214}{{\ttfamily hep-ph/9405214}}].

\bibitem{aqgcold}
O.~Eboli, M.~Gonzalez-Garcia and J.~Mizukoshi, \emph{{$p p \rightarrow j j
  e^{\pm} \mu^{\pm} \nu \nu$ and $j j e^{\pm} \mu^{\mp} \nu \nu$ at
  $\mathcal{O}( \alpha_{\pm}^6)$ and $\mathcal{O}( \alpha_{\pm}^4 \alpha_s^2)$
  for the study of the quartic electroweak gauge boson vertex at CERN LHC}},
  \href{https://doi.org/10.1103/PhysRevD.74.073005}{\emph{Phys. Rev. D}
  {\bfseries 74} (2006) 073005}
  [\href{https://arxiv.org/abs/hep-ph/0606118}{{\ttfamily hep-ph/0606118}}].

\bibitem{aqgcnew}
O.~J.~P. \'Eboli and M.~C. Gonzalez-Garcia, \emph{{Classifying the bosonic
  quartic couplings}},
  \href{https://doi.org/10.1103/PhysRevD.93.093013}{\emph{Phys. Rev. D}
  {\bfseries 93} (2016) 093013}
  [\href{https://arxiv.org/abs/1604.03555}{{\ttfamily 1604.03555}}].

\bibitem{sswwexp1}
{\scshape ATLAS} collaboration, \emph{{Evidence for Electroweak Production of
  $W^{\pm}W^{\pm}jj$ in $pp$ Collisions at $\sqrt{s}=8$ TeV with the ATLAS
  Detector}}, \href{https://doi.org/10.1103/PhysRevLett.113.141803}{\emph{Phys.
  Rev. Lett.} {\bfseries 113} (2014) 141803}
  [\href{https://arxiv.org/abs/1405.6241}{{\ttfamily 1405.6241}}].

\bibitem{sswwexp2}
{\scshape CMS} collaboration, \emph{{Measurements of production cross sections
  of $WZ$ and same-sign $WW$ boson pairs in association with two jets in
  proton-proton collisions at $\sqrt{s} =$ 13 TeV}},
  \href{https://doi.org/10.1016/j.physletb.2020.135710}{\emph{Phys. Lett. B}
  {\bfseries 809} (2020) 135710}
  [\href{https://arxiv.org/abs/2005.01173}{{\ttfamily 2005.01173}}].

\bibitem{zaexp1}
{\scshape ATLAS} collaboration, \emph{{Studies of $Z\gamma$ production in
  association with a high-mass dijet system in $pp$ collisions at $\sqrt{s}=$ 8
  TeV with the ATLAS detector}},
  \href{https://doi.org/10.1007/JHEP07(2017)107}{\emph{JHEP} {\bfseries 07}
  (2017) 107} [\href{https://arxiv.org/abs/1705.01966}{{\ttfamily
  1705.01966}}].

\bibitem{zaexp2}
{\scshape CMS} collaboration, \emph{{Measurement of the cross section for
  electroweak production of $Z \gamma$ in association with two jets and
  constraints on anomalous quartic gauge couplings in proton\textendash{}proton
  collisions at $\sqrt{s} = 8$ TeV}},
  \href{https://doi.org/10.1016/j.physletb.2017.04.071}{\emph{Phys. Lett. B}
  {\bfseries 770} (2017) 380}
  [\href{https://arxiv.org/abs/1702.03025}{{\ttfamily 1702.03025}}].

\bibitem{zaexp3}
{\scshape CMS} collaboration, \emph{{Measurement of the cross section for
  electroweak production of a $Z$ boson, a photon and two jets in proton-proton
  collisions at $\sqrt{s} =$ 13 TeV and constraints on anomalous quartic
  couplings}}, \href{https://doi.org/10.1007/JHEP06(2020)076}{\emph{JHEP}
  {\bfseries 06} (2020) 076}
  [\href{https://arxiv.org/abs/2002.09902}{{\ttfamily 2002.09902}}].

\bibitem{waexp1}
{\scshape CMS} collaboration, \emph{{Measurement of electroweak-induced
  production of W$\gamma$ with two jets in pp collisions at $ \sqrt{s}=8 $ TeV
  and constraints on anomalous quartic gauge couplings}},
  \href{https://doi.org/10.1007/JHEP06(2017)106}{\emph{JHEP} {\bfseries 06}
  (2017) 106} [\href{https://arxiv.org/abs/1612.09256}{{\ttfamily
  1612.09256}}].

\bibitem{zzexp1}
{\scshape CMS} collaboration, \emph{{Measurement of vector boson scattering and
  constraints on anomalous quartic couplings from events with four leptons and
  two jets in proton\textendash{}proton collisions at $\sqrt{s}=$ 13 TeV}},
  \href{https://doi.org/10.1016/j.physletb.2017.10.020}{\emph{Phys. Lett. B}
  {\bfseries 774} (2017) 682}
  [\href{https://arxiv.org/abs/1708.02812}{{\ttfamily 1708.02812}}].

\bibitem{zzexp2}
{\scshape CMS} collaboration, \emph{{Measurement of differential cross sections
  for Z boson pair production in association with jets at $\sqrt{s} =$ 8 and 13
  TeV}}, \href{https://doi.org/10.1016/j.physletb.2018.11.007}{\emph{Phys.
  Lett. B} {\bfseries 789} (2019) 19}
  [\href{https://arxiv.org/abs/1806.11073}{{\ttfamily 1806.11073}}].

\bibitem{wzexp1}
{\scshape ATLAS} collaboration, \emph{{Observation of electroweak $W^{\pm}Z$
  boson pair production in association with two jets in $pp$ collisions at
  $\sqrt{s} =$ 13 TeV with the ATLAS detector}},
  \href{https://doi.org/10.1016/j.physletb.2019.05.012}{\emph{Phys. Lett. B}
  {\bfseries 793} (2019) 469}
  [\href{https://arxiv.org/abs/1812.09740}{{\ttfamily 1812.09740}}].

\bibitem{wzexp2}
{\scshape CMS} collaboration, \emph{{Measurement of electroweak $WZ$ boson
  production and search for new physics in $WZ$ + two jets events in $pp$
  collisions at $\sqrt{s} = 13$ TeV}},
  \href{https://doi.org/10.1016/j.physletb.2019.05.042}{\emph{Phys. Lett. B}
  {\bfseries 795} (2019) 281}
  [\href{https://arxiv.org/abs/1901.04060}{{\ttfamily 1901.04060}}].

\bibitem{wwexp1}
{\scshape CMS} collaboration, \emph{{Evidence for exclusive $\gamma\gamma \to
  W^+ W^-$ production and constraints on anomalous quartic gauge couplings in
  $pp$ collisions at $ \sqrt{s}=7 $ and 8 TeV}},
  \href{https://doi.org/10.1007/JHEP08(2016)119}{\emph{JHEP} {\bfseries 08}
  (2016) 119} [\href{https://arxiv.org/abs/1604.04464}{{\ttfamily
  1604.04464}}].

\bibitem{wwexp2}
{\scshape CMS} collaboration, \emph{{Observation of electroweak production of
  same-sign W boson pairs in the two jet and two same-sign lepton final state
  in proton-proton collisions at $\sqrt{s} = $ 13 TeV}},
  \href{https://doi.org/10.1103/PhysRevLett.120.081801}{\emph{Phys. Rev. Lett.}
  {\bfseries 120} (2018) 081801}
  [\href{https://arxiv.org/abs/1709.05822}{{\ttfamily 1709.05822}}].

\bibitem{wvzvexp}
{\scshape CMS} collaboration, \emph{{Search for anomalous electroweak
  production of vector boson pairs in association with two jets in
  proton-proton collisions at 13 TeV}},
  \href{https://doi.org/10.1016/j.physletb.2019.134985}{\emph{Phys. Lett. B}
  {\bfseries 798} (2019) 134985}
  [\href{https://arxiv.org/abs/1905.07445}{{\ttfamily 1905.07445}}].

\bibitem{waexp2}
{\scshape CMS} collaboration, \emph{{Observation of electroweak production of
  W$\gamma$ with two jets in proton-proton collisions at $\sqrt {s}$ = 13
  TeV}}, \href{https://doi.org/10.1016/j.physletb.2020.135988}{\emph{Phys.
  Lett. B} {\bfseries 811} (2020) 135988}
  [\href{https://arxiv.org/abs/2008.10521}{{\ttfamily 2008.10521}}].

\bibitem{zzexp3}
{\scshape CMS} collaboration, \emph{{Evidence for electroweak production of
  four charged leptons and two jets in proton-proton collisions at $\sqrt {s}$
  = 13 TeV}}, \href{https://doi.org/10.1016/j.physletb.2020.135992}{\emph{Phys.
  Lett. B} {\bfseries 812} (2021) 135992}
  [\href{https://arxiv.org/abs/2008.07013}{{\ttfamily 2008.07013}}].

\bibitem{ntgc1}
C.~Degrande, \emph{{A basis of dimension-eight operators for anomalous neutral
  triple gauge boson interactions}},
  \href{https://doi.org/10.1007/JHEP02(2014)101}{\emph{JHEP} {\bfseries 02}
  (2014) 101} [\href{https://arxiv.org/abs/1308.6323}{{\ttfamily 1308.6323}}].

\bibitem{ntgc2}
J.~Ellis, S.-F. Ge, H.-J. He and R.-Q. Xiao, \emph{{Probing the scale of new
  physics in the $ZZ\gamma$ coupling at $e^+e^-$ colliders}},
  \href{https://doi.org/10.1088/1674-1137/44/6/063106}{\emph{Chin. Phys. C}
  {\bfseries 44} (2020) 063106}
  [\href{https://arxiv.org/abs/1902.06631}{{\ttfamily 1902.06631}}].

\bibitem{ntgc3}
J.~Ellis, H.-J. He and R.-Q. Xiao, \emph{{Probing new physics in dimension-8
  neutral gauge couplings at $e^+e^-$ colliders}},
  \href{https://doi.org/10.1007/s11433-020-1617-3}{\emph{Sci. China Phys. Mech.
  Astron.} {\bfseries 64} (2021) 221062}
  [\href{https://arxiv.org/abs/2008.04298}{{\ttfamily 2008.04298}}].

\bibitem{ntgc4}
G.~J. Gounaris, J.~Layssac and F.~M. Renard, \emph{{Off-shell structure of the
  anomalous $Z$ and $\gamma$ selfcouplings}},
  \href{https://doi.org/10.1103/PhysRevD.65.017302}{\emph{Phys. Rev. D}
  {\bfseries 62} (2000) 073012}
  [\href{https://arxiv.org/abs/hep-ph/0005269}{{\ttfamily hep-ph/0005269}}].

\bibitem{ntgc5}
G.~J. Gounaris, J.~Layssac and F.~M. Renard, \emph{{Signatures of the anomalous
  $Z\gamma$ and $Z Z$ production at the lepton and hadron colliders}},
  \href{https://doi.org/10.1103/PhysRevD.61.073013}{\emph{Phys. Rev. D}
  {\bfseries 61} (2000) 073013}
  [\href{https://arxiv.org/abs/hep-ph/9910395}{{\ttfamily hep-ph/9910395}}].

\bibitem{ntgc6}
A.~Senol, H.~Denizli, A.~Yilmaz, I.~Turk~Cakir, K.~Y. Oyulmaz, O.~Karadeniz
  et~al., \emph{{Probing the Effects of Dimension-eight Operators Describing
  Anomalous Neutral Triple Gauge Boson Interactions at FCC-hh}},
  \href{https://doi.org/10.1016/j.nuclphysb.2018.08.018}{\emph{Nucl. Phys. B}
  {\bfseries 935} (2018) 365}
  [\href{https://arxiv.org/abs/1805.03475}{{\ttfamily 1805.03475}}].

\bibitem{ntgc7}
Q.~Fu, J.-C. Yang, C.-X. Yue and Y.-C. Guo, \emph{{The study of neutral triple
  gauge couplings in the process $e^+e^-\to Z\gamma$ including unitarity
  bounds}}, \href{https://doi.org/10.1016/j.nuclphysb.2021.115543}{\emph{Nucl.
  Phys. B} {\bfseries 972} (2021) 115543}
  [\href{https://arxiv.org/abs/2102.03623}{{\ttfamily 2102.03623}}].

\bibitem{wwstudy}
Y.-C. Guo, Y.-Y. Wang and J.-C. Yang, \emph{{Constraints on anomalous quartic
  gauge couplings by $\gamma\gamma \to W^+W^-$ scattering}},
  \href{https://doi.org/10.1016/j.nuclphysb.2020.115222}{\emph{Nucl. Phys. B}
  {\bfseries 961} (2020) 115222}
  [\href{https://arxiv.org/abs/1912.10686}{{\ttfamily 1912.10686}}].

\bibitem{wastudy}
Y.-C. Guo, Y.-Y. Wang, J.-C. Yang and C.-X. Yue, \emph{{Constraints on
  anomalous quartic gauge couplings via $W\gamma jj$ production at the LHC}},
  \href{https://doi.org/10.1088/1674-1137/abb4d2}{\emph{Chin. Phys. C}
  {\bfseries 44} (2020) 123105}
  [\href{https://arxiv.org/abs/2002.03326}{{\ttfamily 2002.03326}}].

\bibitem{zastudy}
J.-C. Yang, Y.-C. Guo, C.-X. Yue and Q.~Fu, \emph{{Constraints on anomalous
  quartic gauge couplings via $Z\gamma jj$ production at the LHC}},
  \href{https://doi.org/10.1103/PhysRevD.104.035015}{\emph{Phys. Rev. D}
  {\bfseries 104} (2021) 035015}
  [\href{https://arxiv.org/abs/2107.01123}{{\ttfamily 2107.01123}}].

\bibitem{wwwwunitary}
J.-C. Yang, J.-H. Chen and Y.-C. Guo, \emph{{Extract the energy scale of
  anomalous $\gamma\gamma\rightarrow W^+W^-$ scattering in the vector boson
  scattering process using artificial neural networks}},
  \href{https://doi.org/10.1007/JHEP09(2021)085}{\emph{JHEP} {\bfseries 21}
  (2020) 085} [\href{https://arxiv.org/abs/2107.13624}{{\ttfamily
  2107.13624}}].

\bibitem{Yang:2022ilt}
J.-C. Yang, Y.-C. Guo, B.~Liu and T.~Li, \emph{{Shining light on magnetic
  monopoles through high-energy muon colliders}},
  \href{https://doi.org/10.1016/j.nuclphysb.2023.116097}{\emph{Nucl. Phys. B}
  {\bfseries 987} (2023) 116097}
  [\href{https://arxiv.org/abs/2208.02188}{{\ttfamily 2208.02188}}].

\bibitem{Yang:2022fhw}
J.-C. Yang, X.-Y. Han, Z.-B. Qin, T.~Li and Y.-C. Guo, \emph{{Measuring the
  anomalous quartic gauge couplings in the $W^{+}W^-\rightarrow W^+W^-$ process
  at muon collider using artificial neural networks}},
  \href{https://doi.org/10.1007/JHEP09(2022)074}{\emph{JHEP} {\bfseries 09}
  (2022) 074} [\href{https://arxiv.org/abs/2204.10034}{{\ttfamily
  2204.10034}}].

\bibitem{triphoton}
J.-C. Yang, Z.-B. Qing, X.-Y. Han, Y.-C. Guo and T.~Li, \emph{{Tri-photon at
  muon collider: a new process to probe the anomalous quartic gauge
  couplings}}, \href{https://doi.org/10.1007/JHEP07(2022)053}{\emph{JHEP}
  {\bfseries 22} (2020) 053}
  [\href{https://arxiv.org/abs/2204.08195}{{\ttfamily 2204.08195}}].

\bibitem{Zhang:2023yfg}
S.~Zhang, J.-C. Yang and Y.-C. Guo, \emph{{Using k-means assistant event
  selection strategy to study anomalous quartic gauge couplings at muon
  colliders}},  \href{https://arxiv.org/abs/2302.01274}{{\ttfamily
  2302.01274}}.

\bibitem{Ellis:2022zdw}
J.~Ellis, H.-J. He and R.-Q. Xiao, \emph{{Probing Neutral Triple Gauge
  Couplings at the LHC and Future Hadron Colliders}},
  \href{https://arxiv.org/abs/2206.11676}{{\ttfamily 2206.11676}}.

\bibitem{Bachas:1995kx}
C.~Bachas, \emph{{D-brane dynamics}},
  \href{https://doi.org/10.1016/0370-2693(96)00238-9}{\emph{Phys. Lett. B}
  {\bfseries 374} (1996) 37}
  [\href{https://arxiv.org/abs/hep-th/9511043}{{\ttfamily hep-th/9511043}}].

\bibitem{Fradkin:1985qd}
E.~S. Fradkin and A.~A. Tseytlin, \emph{{Nonlinear Electrodynamics from
  Quantized Strings}},
  \href{https://doi.org/10.1016/0370-2693(85)90205-9}{\emph{Phys. Lett. B}
  {\bfseries 163} (1985) 123}.

\bibitem{Tseytlin:1999dj}
A.~A. Tseytlin, \emph{{Born-Infeld action, supersymmetry and string theory}},
  \href{https://arxiv.org/abs/hep-th/9908105}{{\ttfamily hep-th/9908105}}.

\bibitem{Ellis:2021dfa}
J.~Ellis, S.-F. Ge and K.~Ma, \emph{{Hadron collider probes of the quartic
  couplings of gluons to the photon and Z boson}},
  \href{https://doi.org/10.1007/JHEP04(2022)123}{\emph{JHEP} {\bfseries 04}
  (2022) 123} [\href{https://arxiv.org/abs/2112.06729}{{\ttfamily
  2112.06729}}].

\bibitem{Tikhonin:2008pw}
F.~F. Tikhonin, \emph{{On the effects at colliding mu meson beams}},
  \href{https://arxiv.org/abs/0805.3961}{{\ttfamily 0805.3961}}.

\bibitem{Neuffer:1979gq}
D.~Neuffer, \emph{{Colliding Muon Beams at 90 GeV}}, .

\bibitem{Cline:1980sa}
D.~Cline and D.~Neuffer, \emph{{A muon storage ring for neutrino oscillations
  experiments}}, \href{https://doi.org/10.1063/1.2948637}{\emph{AIP Conf.
  Proc.} {\bfseries 68} (1980) 856}.

\bibitem{Skrinsky:1981ht}
A.~N. Skrinsky and V.~V. Parkhomchuk, \emph{{Cooling Methods for Beams of
  Charged Particles. (In Russian)}}, {\emph{Sov. J. Part. Nucl.} {\bfseries 12}
  (1981) 223}.

\bibitem{Neuffer:1983jr}
D.~Neuffer, \emph{{Principles and Applications of Muon Cooling}},
  \href{https://doi.org/10.2172/1156195}{\emph{Part. Accel.} {\bfseries 14}
  (1983) 75}.

\bibitem{Black:2022cth}
K.~M. Black et~al., \emph{{Muon Collider Forum Report}},
  \href{https://arxiv.org/abs/2209.01318}{{\ttfamily 2209.01318}}.

\bibitem{Accettura:2023ked}
C.~Accettura et~al., \emph{{Towards a Muon Collider}},
  \href{https://arxiv.org/abs/2303.08533}{{\ttfamily 2303.08533}}.

\bibitem{Han:2020pif}
T.~Han, D.~Liu, I.~Low and X.~Wang, \emph{{Electroweak couplings of the Higgs
  boson at a multi-TeV muon collider}},
  \href{https://doi.org/10.1103/PhysRevD.103.013002}{\emph{Phys. Rev. D}
  {\bfseries 103} (2021) 013002}
  [\href{https://arxiv.org/abs/2008.12204}{{\ttfamily 2008.12204}}].

\bibitem{Han:2020uak}
T.~Han, Z.~Liu, L.-T. Wang and X.~Wang, \emph{{WIMPs at High Energy Muon
  Colliders}}, \href{https://doi.org/10.1103/PhysRevD.103.075004}{\emph{Phys.
  Rev. D} {\bfseries 103} (2021) 075004}
  [\href{https://arxiv.org/abs/2009.11287}{{\ttfamily 2009.11287}}].

\bibitem{Liu:2021jyc}
W.~Liu and K.-P. Xie, \emph{{Probing electroweak phase transition with
  multi-TeV muon colliders and gravitational waves}},
  \href{https://doi.org/10.1007/JHEP04(2021)015}{\emph{JHEP} {\bfseries 04}
  (2021) 015} [\href{https://arxiv.org/abs/2101.10469}{{\ttfamily
  2101.10469}}].

\bibitem{Liu:2021akf}
W.~Liu, K.-P. Xie and Z.~Yi, \emph{{Testing leptogenesis at the LHC and future
  muon colliders: A Z' scenario}},
  \href{https://doi.org/10.1103/PhysRevD.105.095034}{\emph{Phys. Rev. D}
  {\bfseries 105} (2022) 095034}
  [\href{https://arxiv.org/abs/2109.15087}{{\ttfamily 2109.15087}}].

\bibitem{Han:2021udl}
T.~Han, S.~Li, S.~Su, W.~Su and Y.~Wu, \emph{{Heavy Higgs bosons in 2HDM at a
  muon collider}},
  \href{https://doi.org/10.1103/PhysRevD.104.055029}{\emph{Phys. Rev. D}
  {\bfseries 104} (2021) 055029}
  [\href{https://arxiv.org/abs/2102.08386}{{\ttfamily 2102.08386}}].

\bibitem{Han:2021lnp}
T.~Han, W.~Kilian, N.~Kreher, Y.~Ma, J.~Reuter, T.~Striegl et~al.,
  \emph{{Precision test of the muon-Higgs coupling at a high-energy muon
  collider}}, \href{https://doi.org/10.1007/JHEP12(2021)162}{\emph{JHEP}
  {\bfseries 12} (2021) 162}
  [\href{https://arxiv.org/abs/2108.05362}{{\ttfamily 2108.05362}}].

\bibitem{Han:2022mzp}
T.~Han, T.~Li and X.~Wang, \emph{{Axion-Like Particles at High Energy Muon
  Colliders -- A White paper for Snowmass 2021}},  in \emph{{Snowmass 2021}},
  3, 2022, \href{https://arxiv.org/abs/2203.05484}{{\ttfamily 2203.05484}}.

\bibitem{Han:2022ubw}
T.~Han, Z.~Liu, L.-T. Wang and X.~Wang, \emph{{WIMP Dark Matter at High Energy
  Muon Colliders $-$A White Paper for Snowmass 2021}},  in \emph{{Snowmass
  2021}}, 3, 2022, \href{https://arxiv.org/abs/2203.07351}{{\ttfamily
  2203.07351}}.

\bibitem{Han:2022edd}
T.~Han, S.~Li, S.~Su, W.~Su and Y.~Wu, \emph{{BSM Higgs Production at a Muon
  Collider}},  in \emph{{Snowmass 2021}}, 5, 2022,
  \href{https://arxiv.org/abs/2205.11730}{{\ttfamily 2205.11730}}.

\bibitem{Chowdhury:2023imd}
T.~A. Chowdhury, A.~Jueid, S.~Nasri and S.~Saad, \emph{{Probing Zee-Babu states
  at Muon Colliders}},  \href{https://arxiv.org/abs/2306.01255}{{\ttfamily
  2306.01255}}.

\bibitem{Jueid:2023zxx}
A.~Jueid and S.~Nasri, \emph{{Lepton portal dark matter at muon colliders:
  Total rates and generic features for phenomenologically viable scenarios}},
  \href{https://doi.org/10.1103/PhysRevD.107.115027}{\emph{Phys. Rev. D}
  {\bfseries 107} (2023) 115027}
  [\href{https://arxiv.org/abs/2301.12524}{{\ttfamily 2301.12524}}].

\bibitem{Aime:2022flm}
C.~Aime et~al., \emph{{Muon Collider Physics Summary}},
  \href{https://arxiv.org/abs/2203.07256}{{\ttfamily 2203.07256}}.

\bibitem{Buchmuller:1985jz}
W.~Buchmuller and D.~Wyler, \emph{{Effective Lagrangian Analysis of New
  Interactions and Flavor Conservation}},
  \href{https://doi.org/10.1016/0550-3213(86)90262-2}{\emph{Nucl. Phys. B}
  {\bfseries 268} (1986) 621}.

\bibitem{AlAli:2021let}
H.~Al~Ali et~al., \emph{{The muon Smasher\textquoteright{}s guide}},
  \href{https://doi.org/10.1088/1361-6633/ac6678}{\emph{Rept. Prog. Phys.}
  {\bfseries 85} (2022) 084201}
  [\href{https://arxiv.org/abs/2103.14043}{{\ttfamily 2103.14043}}].

\bibitem{eva1}
G.~L. Kane, W.~W. Repko and W.~B. Rolnick, \emph{{The Effective W+-, Z0
  Approximation for High-Energy Collisions}},
  \href{https://doi.org/10.1016/0370-2693(84)90105-9}{\emph{Phys. Lett. B}
  {\bfseries 148} (1984) 367}.

\bibitem{eva2}
E.~Boos, H.~J. He, W.~Kilian, A.~Pukhov, C.~P. Yuan and P.~M. Zerwas,
  \emph{{Strongly interacting vector bosons at TeV e+ e- linear colliders}},
  \href{https://doi.org/10.1103/PhysRevD.57.1553}{\emph{Phys. Rev. D}
  {\bfseries 57} (1998) 1553}
  [\href{https://arxiv.org/abs/hep-ph/9708310}{{\ttfamily hep-ph/9708310}}].

\bibitem{eva3}
R.~Ruiz, A.~Costantini, F.~Maltoni and O.~Mattelaer, \emph{{The Effective
  Vector Boson Approximation in High-Energy Muon Collisions}},  11, 2021.

\bibitem{unitarityHistory1}
T.~Lee and C.-N. Yang, \emph{{THEORETICAL DISCUSSIONS ON POSSIBLE HIGH-ENERGY
  NEUTRINO EXPERIMENTS}},
  \href{https://doi.org/10.1103/PhysRevLett.4.307}{\emph{Phys. Rev. Lett.}
  {\bfseries 4} (1960) 307}.

\bibitem{unitarityHistory2}
M.~Froissart, \emph{{Asymptotic behavior and subtractions in the Mandelstam
  representation}}, \href{https://doi.org/10.1103/PhysRev.123.1053}{\emph{Phys.
  Rev.} {\bfseries 123} (1961) 1053}.

\bibitem{unitarityHistory3}
G.~Passarino, \emph{{W W scattering and perturbative unitarity}},
  \href{https://doi.org/10.1016/0550-3213(90)90593-3}{\emph{Nucl. Phys. B}
  {\bfseries 343} (1990) 31}.

\bibitem{partialwaveunitaritybound}
T.~Corbett, O.~J.~P. \'Eboli and M.~C. Gonzalez-Garcia, \emph{{Unitarity
  Constraints on Dimension-Six Operators}},
  \href{https://doi.org/10.1103/PhysRevD.91.035014}{\emph{Phys. Rev. D}
  {\bfseries 91} (2015) 035014}
  [\href{https://arxiv.org/abs/1411.5026}{{\ttfamily 1411.5026}}].

\bibitem{ubnew1}
E.~d.~S. Almeida, O.~J.~P. \'Eboli and M.~C. Gonzalez\textendash{}Garcia,
  \emph{{Unitarity constraints on anomalous quartic couplings}},
  \href{https://doi.org/10.1103/PhysRevD.101.113003}{\emph{Phys. Rev. D}
  {\bfseries 101} (2020) 113003}
  [\href{https://arxiv.org/abs/2004.05174}{{\ttfamily 2004.05174}}].

\bibitem{ubnew2}
W.~Kilian, S.~Sun, Q.-S. Yan, X.~Zhao and Z.~Zhao, \emph{{Multi-Higgs boson
  production and unitarity in vector-boson fusion at future hadron colliders}},
  \href{https://doi.org/10.1103/PhysRevD.101.076012}{\emph{Phys. Rev. D}
  {\bfseries 101} (2020) 076012}
  [\href{https://arxiv.org/abs/1808.05534}{{\ttfamily 1808.05534}}].

\bibitem{partialwaveexpansion}
M.~Jacob and G.~Wick, \emph{{On the General Theory of Collisions for Particles
  with Spin}}, \href{https://doi.org/10.1016/0003-4916(59)90051-X}{\emph{Annals
  Phys.} {\bfseries 7} (1959) 404}.

\bibitem{unitarity1}
J.~Layssac, F.~Renard and G.~Gounaris, \emph{{Unitarity constraints for
  transverse gauge bosons at LEP and supercolliders}},
  \href{https://doi.org/10.1016/0370-2693(94)90872-9}{\emph{Phys. Lett. B}
  {\bfseries 332} (1994) 146}
  [\href{https://arxiv.org/abs/hep-ph/9311370}{{\ttfamily hep-ph/9311370}}].

\bibitem{unitarity2}
T.~Corbett, O.~\'Eboli and M.~Gonzalez-Garcia, \emph{{Unitarity Constraints on
  Dimension-six Operators II: Including Fermionic Operators}},
  \href{https://doi.org/10.1103/PhysRevD.96.035006}{\emph{Phys. Rev. D}
  {\bfseries 96} (2017) 035006}
  [\href{https://arxiv.org/abs/1705.09294}{{\ttfamily 1705.09294}}].

\bibitem{unitarity3}
R.~Gomez-Ambrosio, \emph{{Vector Boson Scattering Studies in CMS: The $pp \to
  ZZ jj$ Channel}},
  \href{https://doi.org/10.5506/APhysPolBSupp.11.239}{\emph{Acta Phys. Polon.
  Supp.} {\bfseries 11} (2018) 239}
  [\href{https://arxiv.org/abs/1807.09634}{{\ttfamily 1807.09634}}].

\bibitem{unitarity4}
G.~Perez, M.~Sekulla and D.~Zeppenfeld, \emph{{Anomalous quartic gauge
  couplings and unitarization for the vector boson scattering process
  $pp\rightarrow W^+W^+jjX\rightarrow \ell ^+\nu _\ell \ell ^+\nu _\ell jjX$}},
  \href{https://doi.org/10.1140/epjc/s10052-018-6230-1}{\emph{Eur. Phys. J. C}
  {\bfseries 78} (2018) 759}
  [\href{https://arxiv.org/abs/1807.02707}{{\ttfamily 1807.02707}}].

\bibitem{madgraph}
J.~Alwall, R.~Frederix, S.~Frixione, V.~Hirschi, F.~Maltoni, O.~Mattelaer
  et~al., \emph{{The automated computation of tree-level and next-to-leading
  order differential cross sections, and their matching to parton shower
  simulations}}, \href{https://doi.org/10.1007/JHEP07(2014)079}{\emph{JHEP}
  {\bfseries 07} (2014) 079} [\href{https://arxiv.org/abs/1405.0301}{{\ttfamily
  1405.0301}}].

\bibitem{feynrules}
N.~D. Christensen and C.~Duhr, \emph{{FeynRules - Feynman rules made easy}},
  \href{https://doi.org/10.1016/j.cpc.2009.02.018}{\emph{Comput. Phys. Commun.}
  {\bfseries 180} (2009) 1614}
  [\href{https://arxiv.org/abs/0806.4194}{{\ttfamily 0806.4194}}].

\bibitem{pythia}
T.~Sj\"ostrand, S.~Ask, J.~R. Christiansen, R.~Corke, N.~Desai, P.~Ilten
  et~al., \emph{{An introduction to PYTHIA 8.2}},
  \href{https://doi.org/10.1016/j.cpc.2015.01.024}{\emph{Comput. Phys. Commun.}
  {\bfseries 191} (2015) 159}
  [\href{https://arxiv.org/abs/1410.3012}{{\ttfamily 1410.3012}}].

\bibitem{NNPDF}
{\scshape NNPDF} collaboration, \emph{{Parton distributions with QED
  corrections}},
  \href{https://doi.org/10.1016/j.nuclphysb.2013.10.010}{\emph{Nucl. Phys. B}
  {\bfseries 877} (2013) 290}
  [\href{https://arxiv.org/abs/1308.0598}{{\ttfamily 1308.0598}}].

\bibitem{delphes}
{\scshape DELPHES 3} collaboration, \emph{{DELPHES 3, A modular framework for
  fast simulation of a generic collider experiment}},
  \href{https://doi.org/10.1007/JHEP02(2014)057}{\emph{JHEP} {\bfseries 02}
  (2014) 057} [\href{https://arxiv.org/abs/1307.6346}{{\ttfamily 1307.6346}}].

\bibitem{Guo:2023nfu}
Y.-C. Guo, F.~Feng, A.~Di, S.-Q. Lu and J.-C. Yang, \emph{{MLAnalysis: An
  open-source program for high energy physics analyses}},
  \href{https://doi.org/10.1016/j.cpc.2023.108957}{\emph{Comput. Phys. Commun.}
  {\bfseries 294} (2024) 108957}
  [\href{https://arxiv.org/abs/2305.00964}{{\ttfamily 2305.00964}}].

\bibitem{Cowan:2010js}
G.~Cowan, K.~Cranmer, E.~Gross and O.~Vitells, \emph{{Asymptotic formulae for
  likelihood-based tests of new physics}},
  \href{https://doi.org/10.1140/epjc/s10052-011-1554-0}{\emph{Eur. Phys. J. C}
  {\bfseries 71} (2011) 1554}
  [\href{https://arxiv.org/abs/1007.1727}{{\ttfamily 1007.1727}}].

\bibitem{pdgss:2020}
{\scshape Particle Data Group} collaboration, \emph{{Review of Particle
  Physics}}, \href{https://doi.org/10.1093/ptep/ptaa104}{\emph{PTEP} {\bfseries
  2020} (2020) 083C01}.

\bibitem{Rauch:2016pai}
M.~Rauch, \emph{{Vector-Boson Fusion and Vector-Boson Scattering}},
  \href{https://arxiv.org/abs/1610.08420}{{\ttfamily 1610.08420}}.

\end{thebibliography}\endgroup
\bibliographystyle{JHEP}

\end{document}